\documentclass[namedreferences,hyperref,optionalrh,solaromanenum]{spr-sola}

\usepackage{amsmath}
\usepackage{graphicx}                    % For eps figures, newer & more powerfull
\usepackage{color}                       % For color text: \color command
\begin{document}

\begin{frontmatter}

\title{Solar Wind Clustering at Mars using Machine Learning Techniques}

%%%%%%%%%%%%%%%%%%%%%%%%%%%%%%%%%%%%%%%%%%%%%%%%%%%
%% Authors Names
%

\author[addressref={1},corref,email={silvia.ferro@kuleuven.be}]{\inits{S. }\fnm{Silvia } \snm{Ferro}\orcid{0009-0001-2448-9854}}
\author[addressref={2},corref,email={catherine.regan@mail.wvu.edu}]{\inits{C.E. }\fnm{Catherine E. }\snm{Regan}\orcid{0000-0001-8784-5919}}
\author[addressref={3},corref,email={austin.smith@lasp.colorado.edu}]{\inits{A. M. }\fnm{Austin M. }\snm{Smith}\orcid{0009-0001-5822-9735}}
\author[addressref={4} ,corref,email={alvin.angeles@unh.edu}]{\inits{A.J.G. }\fnm{Alvin J. G. }\snm{Angeles}\orcid{0009-0006-6493-5851}}
\author[addressref={5}]{\inits{N.A. }\fnm{Nicholas A. }\snm{Gross}}
\author[addressref={6}]{\inits{F. }\fnm{Farzad }\snm{Kamalabadi}}
\author[addressref={7}]{\inits{M. }\fnm{Marco }\snm{Velli}}
\author[addressref={8}]{\inits{J.S.}\fnm{Jasper S. }\snm{Halekas}}

%%%%%%%%%%%%%%%%%%%%%%%%%%%%%%%%%%%%%%%%%%%%%%%%%%%
%% Affilations 
%% id shold be the same with \author addressref value.
\address[id={1}]{Centre for mathematical Plasma Astrophysics, Department of Mathematics, KU Leuven, Celestijnenlaan 200B, B-3001 Leuven, Belgium}
\address[id={2}]{Department of Physics and Astronomy, West Virginia University, Morgantown, WV, USA}
\address[id={3}]{Laboratory for Atmospheric and Space Physics, Smead Aerospace Engineering Sciences Department, University of Colorado Boulder, CO, USA}
\address[id={4}]{Department of Physics and Astronomy, Institute for the Study of Earth, Oceans, and Space, University of New Hampshire, Durham, NH, USA}
\address[id={5}]{Department of Astronomy, Boston University, Boston, MA, USA}
\address[id={6}]{Department of Electrical and Computer Engineering, University of Illinois at Urbana-Champaign, Urbana, IL, USA}
\address[id={7}]{Department of Earth, Planetary and Space Science, California Institute of Technology, Pasadena, CA, USA}
\address[id={8}]{Department of Physics and Astronomy, University of Iowa, Iowa City, IA, USA}
%%%%%%%%%%%%%%%%%%%%%%%%%%%%%%%%%%%%%%%%%%%%%%%%%%%
%%% Abstract 
% \begin{abstract}
%     It is important to understand solar wind conditions throughout the solar system, particularly with the push for human exploration of the Moon and Mars. In order to categorize different solar wind conditions, many parameters need to be taken into account, including densities, velocities, and availability of data in order to gather data. MAVEN has been orbiting Mars for a full solar cycle, and has solar wind data for a large percentage of this. Machine learning techniques allow this data to be normalized and segregated, leading to classifications of different conditions. 
% \end{abstract}
\begin{abstract}
Understanding solar wind variability throughout the heliosphere is essential for fundamental space physics and future exploration of the Moon and Mars. The Mars Atmosphere and Volatile EvolutioN (MAVEN) spacecraft has provided upstream solar wind measurements at Mars spanning Solar Cycles~24 and~25, enabling a statistical investigation of solar wind regimes at this heliocentric distance. In this work, we apply an unsupervised machine-learning framework combining Principal Component Analysis and K-Means clustering to a normalized, multi-dimensional solar wind dataset to identify recurrent solar wind regimes in a physically interpretable, data-driven manner. The resulting clustering identifies representative slow, fast, intermediate, and compressed solar wind regimes whose relative occurrence and temporal organization vary with solar activity. This manuscript is part of the Heliophysics Summer School Machine Learning Special Collection.
\end{abstract}

%%%%%%%%%%%%%%%%%%%%%%%%%%%%%%%%%%%%%%%%%%%%%%%%%%%
%% Keywords
%
%\keywords{}

\end{frontmatter}
%-------------------------------------------------

%%%%%%%%%%%%%%%%%%%%%%%%%%%%%%%%%%%%%%%%%%%%%%%%%%%
%% Sections

\section{Introduction}\label{intro}
Understanding the conditions of the solar wind is of increasing importance with the drive for human exploration of the Moon and Mars \citep{m2mwhitepaper25}. The change in the solar wind on a temporal and spatial scale is variable, changing over the solar cycle in addition to individual space weather events such as solar flares or co-rotating interaction regions (CIRs). Mars does not have an intrinsic magnetic field, but rather experiences a 'hybrid' magnetosphere where it has the combination of both an induced magnetosphere due to the interaction of solar wind with the ionosphere, and an intrinsic magnetosphere in the form of mini-magnetospheres over crustal magnetic field regions \citep{connerney2001}, particularly in the southern hemisphere. This unique environment allows the influence of the solar wind to be seen in the ionosphere on the scale of minutes later. With human exploration of Mars being the goal of the next decade \citep{m2mwhitepaper24} it is a priority to understand this relationship between the solar wind and Mars in order to keep astronauts and equipment safe. Machine learning allows current large datasets to be analyzed through different methodologies that are a major driving influence for scientific discovery. This study utilizes these methods to investigate how the solar wind can be classified at Mars using over 10 years of data from the Mars Atmosphere and Volatile EvolutioN (MAVEN) mission.

Through the use of an unsupervised machine-learning technique which combined Principal Component Analysis (PCA) and K-Means clustering, we analyzed a multi-year dataset from the MAVEN spacecraft spanning Solar Cycles 24 and 25. Our study identified six recurring clusters corresponding to representative solar wind regimes throughout the analyzed interval. We found that their relative appearance, persistence, and organization were heavily influenced by solar activity. 
Conditions at solar minimum were characterized by extended slow solar wind intervals with reduced variability and infrequent compressed states. Conversely, at solar maximum, we observed rapid transitions between plasma regimes, with fast and compressed regimes becoming more prevalent, highlighting the increased complexity of the heliospheric plasma environment. Although each principal component incorporates multiple physical parameters, the combined analysis of PCA loadings, cluster-averaged plasma characteristics, and temporal variability indicates that the resultant groups constitute a physically coherent regime identification of the solar wind at Mars.

%%%%%%%%%%%%%%%%%%%%%%%%%%%%%%%%%%%%%%%%%%%%%%%%%%%
\subsection{Solar Wind at Mars}\label{sw_mars_intro}
The solar wind is often characterized as a stream of energetic particles that continuously radiate away from the Sun and travel through the Interplanetary Medium (IPM). The mechanism for this outflow was first addressed by Eugene Parker in his pioneering 1958 paper \citep{parker1958}. 

%The solar wind is often delineated into two categories: fast and slow wind \citep{feldman2005a}; although it is still debated whether the wind can be truly separated into only two distinct source regions (see \cite{cranmer2017} for a review and a comprehensive list of references). Contemporary studies point to the fast solar wind originating primarily in the central regions of coronal holes, which are predominantly located in the solar poles \citep{bame1993,mccomas2002,zirker1977}. The slow solar wind's origin is more ambiguous. Observations have shown that the slow solar wind is concentrated around the heliospheric current sheet during solar minimum, but its exact origin is still an open question \citep{abbo2016,wang2000,wang2009}. 

The solar wind has been extensively studied by multiple terrestrial and heliospheric missions over the last few decades, with the majority of solar wind statistical studies being conducted via in situ and remote sensing observations. Although these measurements have established the foundation of our current understanding, they primarily characterize the solar wind within 1 AU. Due to the dynamic nature of the solar wind, plasma and magnetic structures evolve with heliospheric distance. Additional sampling beyond this region may shed further insight into how the solar wind develops as it propagates through the interplanetary medium. Mars, orbiting at approximately 1.5 AU, provides a complementary vantage point for characterizing and classifying solar wind populations. 

Average solar wind conditions at Mars have a density of 1.4 $\times$ 10$^{-6}$ m$^{-3}$, velocity of 368.9 km s$^{-1}$ and an interplanetary magnetic field strength of 1.87 nT \citep{liu2021}. The martian magnetosphere provides a unique environment (referred to as 'hybrid') that combines an induced magnetosphere (e.g. Venus, Titan, comets), with an intrinsic magnetosphere (e.g. Earth, Saturn). The solar winds interaction with Mars creates an induced magnetosphere, whilst magnetic anomalies provide an instrinsic component. This system is complex, with changes occurring over a variation of timescales (minutes to years \citep{bertucci2011, Halekas2017b,hall2016,hall2019,edberg2009} and references within). Remnants of an intrinsic magnetosphere remain in some areas of the martian crust, with the strongest being present in the southern hemisphere. This creates 'mini-intrinsic-magnetospheres' that push the magnetosphere further from the surface, causing an ionospheric bulge, impacting magnetic boundaries \citep{edberg2008,gruesbeck2018,garnier2022a,garnier2022b}.

%.  The magnetic fields in the southern hemisphere cause an asymmetry in the magnetosphere, pushing the bow shock further from the surface and creating a bulge in the ionosphere. This shifts the induced magnetospheric boundary (IMB), or magnetic pileup boundary (MPB) to higher altitudes over these areas \citep{edberg2008,gruesbeck2018,garnier2022a,garnier2022b}. Average solar wind conditions at Mars have a density of 1.4 $\times$ 10$^{-6}$ m$^{-3}$, velocity of 368.9 km s$^{-1}$ and an interplanetary magnetic field strength of 1.87 nT \citep{liu2021}. \par 

%a martian day (24.7 hours, 1 sol), year (687 days) and with the solar cycle (11 years) (\cite{bertucci2011, Halekas2017b,hall2016,hall2019,edberg2009} and references within).

%Over a martian year ( the bow shock, marking the pressure balance between the solar wind and the magnetosheath, is located approximately 2000 - 5400 km from the surface of Mars at the subsolar and terminator point and changes position by up to 11\% over a martian year \citep{hall2019}. 

The occurrence of dust storm season during southern hemispheric summer can also change magnetospheric dynamics. During global scale dust storms, where the entire planet is covered in a storm, can cause the southern hemispheres bulge to switch it's normal behavior, and dip over crustal fields, adding extra complexity to the system \citep{regan2024,regan2025}. The impact of solar events is also prominent at Mars. In December 2022 MAVEN observed a 'disappearing solar wind' where a significantly low density event caused the ionosphere to expand and the boundaries to decouple \citep{shaver2024}. In addition, during CIRs the magnetic field and solar wind density can increase in values higher than those observed at Earth during the same event \citep{henderson2025}. Understanding of the sensitivity and complexity of the martian magnetosphere is a continuously growing research area, driven by the occurrence of space weather events observed at multiple planets, and the push for future exploration at Mars.  \par

%%%%%%%%%%%%%%%%%%%%%%%%%%%%%%%%%%%%%%%%%%%%%%%%%%%
\subsection{Machine Learning}\label{ml_intro}
Heliospheric science has expanded rapidly in recent years. The 21st century alone has seen a suite of space missions that study everything from the solar corona to the heliopause. These missions have generated a massive and growing volume of data available for analysis which pose challenges for researchers. As highlighted in the review by \cite{asensioramos2023}, the sheer quantity of data that scientists now face makes identifying viable patterns a daunting task. A possible solution has emerged in the field of Machine Learning (ML).

ML is a branch of artificial intelligence in which computer systems learn from data and improve their performance through iterative training. The consistent and repetitive nature of ML algorithms makes them well suited to identifying correlations in massive datasets that would otherwise elude human scrutiny. Space physicists have adopted a variety of ML models to address a myriad of scientific questions. These models usually take on one of three configurations: supervised, unsupervised, and reinforcement learning algorithms \citep{asensioramos2023}.

In supervised learning, a model is trained using samples with known clusterings, allowing it to learn a mapping of inputs to correct outputs. A supervised learning model may, for example, link certain identifiable plasma parameters to fast or slow solar wind streams in training data and then proceed to identify and utilize the trained algorithm to classify large intervals of pristine solar wind. 

Unsupervised learning models are given a dataset and seek to find patterns or connections without prior knowledge of what the "correct" output should be. The strength of unsupervised learning lies in its ability to sift through large volumes of data and identify previously hidden relationships between data points. 

Reinforcement learning borrows techniques from both supervised and unsupervised learning processes but relies on an algorithm learning through interaction with an environment. In reinforcement learning, a model learns through a trial and error process where it takes actions, receives feedback in the form of rewards or penalties, and adjusts its actions to achieve optimization by maximizing its rewards.

Recently, ML has been increasingly utilized in martian studies. Applications range from the use of ML techniques to study Mars' weather patterns, \citep{priyadarshini2021} to investigations into landscape evolution \citep{rajaneesh2022} as well as crustal field morphology \citep{azari2023}. One area that has so far been unexplored is the large scale categorization of solar wind at Mars. Currently, a continuous solar wind dataset is unavailable at Mars due to orbital constraints, with many models propagating conditions at Earth to Mars including machine learning models such as \cite{azari2024} to fill this gap. \cite{azari2024} use a Gaussian regression process to estimate solar wind conditions with highly accurate results when compared to MAVEN measurements within two days. 

The planet is in a unique spot for solar wind monitoring. At approximately 1.5 AU, understanding plasma conditions at Mars would greatly enhance our network of solar wind sensors in the inner heliosphere. Our study expands upon the solar wind clustering framework developed by \cite{amaya2020} and adapts it to the martian plasma environment.

The paper contains the following; Section 2 outlines the MAVEN dataset used and the machine learning techniques, Section 3 contains our results and interpretation and our analysis is concluded in Section 4.

%%%%%%%%%%%%%%%%%%%%%%%%%%%%%%%%%%%%%%%%%%%%%%%%%%%
\section{Methods}\label{methods}
This study utilizes data from the MAVEN mission that has been orbiting Mars since 2014 (\cite{jakosky2017}), covering more than a full solar cycle (solar cycle 24 to 25). MAVEN orbits Mars with a period of approximately 4 hours, sampling regions from the solar wind down to the ionosphere, with a varying elliptical orbit. The spacecraft samples the solar wind a substantial fraction of the time, producing a large dataset of upstream solar wind measurements.

For this study, we use the publicly available synthesized upstream driver dataset provided by Halekas and colleagues, which combines Level 2 measurements from the Solar Wind Ion Analyzer (SWIA) and Magnetometer (MAG) instruments (\cite{Connerney2015,Halekas2015a,Halekas2015b,Halekas2017a,Halekas2017b}; available at \url{https://homepage.physics.uiowa.edu/~jhalekas/drivers.html}). 

%This dataset includes both orbit-averaged and high-resolution (300 points per orbit) data in ASCII format, with columns corresponding to universal time (UT), proton and alpha particle number densities ($n_p$, $n_\alpha$), proton bulk velocity and components ($|v_p|$, $v_x$, $v_y$, $v_z$), proton temperature ($T_p$), and magnetic field components ($B_x$, $B_y$, $B_z$) in Mars Solar Orbital (MSO) coordinates. 

The dataset analyzed in this work therefore provides a robust, physically grounded basis for characterizing the upstream solar wind conditions at Mars, which we further process as described below.

\subsection{MAVEN Dataset and Data Normalization}\label{methods_datanorm}

%\begin{figure}[htbp]
 %   \centering
  %  \includegraphics[width=\linewidth]{Figures/params_cycle24_min.png}
   % \caption{Representative multi-panel time series of solar wind parameters during the minimum of Solar Cycle 24. Panels show (from top to bottom): magnetic field magnitude $|B|$, proton velocity $v_p$, proton density $n_p$, proton temperature $T_p$, and Alfvén velocity $V_A$. Data correspond to the interval 1 October 2018 -- 15 January 2019; they illustrate typical behavior and do not show the full dataset.}
    %\label{fig:cycle24_min_nocluster}
%\end{figure}

The MAVEN dataset analyzed in this study comprises plasma and magnetic field parameters characterized by heterogeneous physical units and markedly different numerical ranges, including proton and alpha particle number densities ($n_p$, $n_\alpha$), solar wind bulk and component velocities ($v_p$, $v_x$, $v_y$, $v_z$), proton temperature ($T_p$), and magnetic field components ($B_x$, $B_y$, $B_z$), as well as several derived quantities. These include the magnetic field magnitude, $|B| = \sqrt{B_x^2 + B_y^2 + B_z^2}$, the proton entropy ratio, $S_p = T_p n_p^{-2/3}$, the Alfvén speed, $V_A = |B| / \sqrt{\mu_0 m_p n_p}$, where $\mu_0 = 4\pi \times 10^{-7},\mathrm{H,m^{-1}}$ is the magnetic permeability of free space and $m_p$ is the proton mass, and a temperature–velocity ratio defined as $T_{\mathrm{ratio}} = (v_p / 258)^{3.113} / T_p$. The 14 data parameters or features are outlined in \cite{Halekas2015b,Halekas2017a}.

%\begin{table}[h]
%\caption{The 14 original data features from the MAVEN database.}
%\label{tab:datafeat}
%\begin{tabular}{c|c|c}
%\hline
%\textbf{Number} & \textbf{Feature} & \textbf{Description} \\
%\hline
%1  & $n_p$ & proton number density \\
%2  & $n_\alpha$ & alpha particle number density \\
%3  & $v_p$ & solar wind bulk velocity \\
%4  & $v_x$ & solar wind x-component\\
%5  & $v_y$ & solar wind y-component\\
%6  & $v_z$ & solar wind z-component\\
%7  & $T_p$ & proton temperature \\
%8  & $B_x$ & magnetic field x-component \\
%9  & $B_y$ & magnetic field y-component \\
%10  & $B_z$ & magnetic field z-component \\
%11 & $\left| B \right|$ & magnetic field magnitude\\
%12 & $S_p$ & proton entropy ratio \\
%13 & $V_A$ & Alfvén speed \\
%14 & $T_{\mathrm{ratio}}$ & temperature ratio \\

%\hline
%\end{tabular}
%\end{table}

%For illustration, \autoref{fig:cycle24_min_nocluster} shows a representative multi-feature time series during the minimum of Solar Cycle~24 (from 1 October 2018 to 15 January 2019), highlighting typical variations in the magnetic field magnitude ($|B|$), proton velocity ($v_p$), proton density ($n_p$), proton temperature ($T_p$), and Alfvén velocity ($V_A$). These plots \textbf{highlight the} variability \textbf{of these parameters at Mars}. 

All input variables were normalized using \emph{z-score normalization}. This normalization was applied on a \emph{feature-wise} basis, ensuring that each variable has zero mean and unit variance prior to further analysis, to avoid individual features disproportionately influencing the dimensionality reduction and clustering procedures.

%For each feature $x$, the normalized value $x_{\mathrm{norm}}$ was computed as
%\begin{equation}
%x_{\mathrm{norm}} = \frac{x - \mu}{\sigma},
%\end{equation}
where $\mu$ and $\sigma$ denote the mean and standard deviation of the feature calculated over the full dataset. 

Normalization was performed before the application of PCA to ensure that the resulting principal components reflect correlations among physical variables rather than differences in absolute scale or magnitude. This pre-processing step allows features such as $n_p$, $v_p$, and $|B|$ to contribute comparably to the dataset's variance structure.

By placing all variables on a common statistical footing, the normalization procedure improves the numerical stability, robustness, and physical interpretability of both the PCA transformation and the subsequent K-Means clustering results.

\subsection{Dimensionality Reduction}\label{methods_dimred}
The original dataset comprises fourteen plasma and magnetic field features, many of which exhibit significant correlations arising from the coupled nature of solar wind plasma properties. The resulting high dimensionality complicates both visualization and statistical analysis, and may increase the influence of noise in subsequent clustering procedures. To address these issues, PCA was employed as a linear, physically interpretable dimensionality-reduction technique.

\begin{figure}[htbp]
    \centering
    \includegraphics[width=\linewidth]{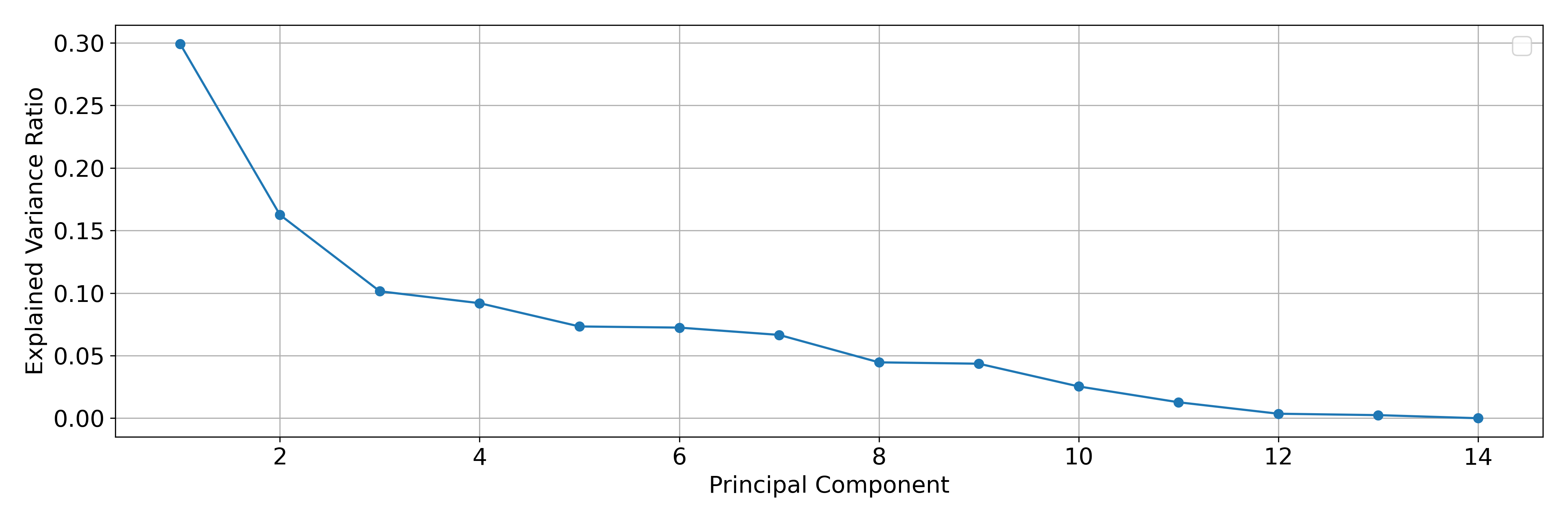}
    \caption{Explained variance ratio of the principal components computed from the normalised MAVEN dataset. The first six components account for the majority of the total variance and are therefore used as the reduced feature space for K-Means clustering.}
    \label{fig:pca}
\end{figure}

PCA was applied to the normalized dataset described in \autoref{methods_datanorm}. The explained variance ratio for each principal component was computed to quantify the contribution of each component to the total variance of the dataset. The results in \autoref{fig:pca} indicate that the first six principal components account for the majority of the total variance (80.1\% in total), thereby justifying a reduction of the original fourteen-dimensional feature space to a six-dimensional representation. 

Several of the input features used in this analysis are not strictly independent, as they include derived quantities that depend nonlinearly on the primary plasma variables, such as the Alfvén speed, proton entropy ratio, and the temperature–velocity ratio. While Principal Component Analysis is a linear technique, the inclusion of these derived features is motivated by physical interpretability rather than statistical independence. \cite{xu2015} defined these partitions to help categorize the solar wind  at 1 AU into four groups; coronal-hole-orgin plasma, streamer-belt-origin plasma, sector-reversal-region plasma and ejecta. Although these clusters were made at Earth, it has been applied on over 50 years of data and is found to be extremely accurate. The presence of these parameters may introduce correlated variance into the PCA space; however, this does not invalidate the analysis, as PCA is used here primarily as a dimensionality-reduction and noise-suppression tool rather than as a strict physical decomposition. In this context, the resulting principal components represent dominant modes of coupled plasma variability rather than orthogonal physical processes.

Consequently, each principal component corresponds to a linear combination of the original physical variables and reflects correlated variations among multiple plasma parameters rather than the behavior of individual quantities in isolation. In particular, the leading components capture coupled plasma properties such as relationships between particle densities and flow velocities, as well as correlations between magnetic field strength and Alfvén speed. Retaining the first six principal components, therefore, provides an effective compromise between information preservation and dimensionality reduction, minimizing information loss while suppressing noise and redundancy in the data.

This reduced PCA space enables clear visualization of the dataset's underlying structure and facilitates more robust clustering. All subsequent clustering analyses presented in this work are performed in the reduced principal component space.

\subsection{K-Means Clustering and Sensitivity Analysis}\label{methods_kmean}

Once in the reduced six-dimensional PCA space, K-Means clustering was applied to identify potentially interesting groupings of the dataset within the feature space. K-Means is an unsupervised learning algorithm that partitions a dataset into a predefined number of clusters based on similarity. The algorithm begins by initializing $k$ cluster centroids and iteratively assigning each data point to the nearest centroid, here this was defined by the Euclidean distance. The centroids are then updated to the mean of the points assigned to each cluster, and this process repeats until convergence, defined here as centroid displacement falling below a tolerance of $10^{-4}$. Ultimately K-Means clustering minimizes the within-cluster sum of squared distances, producing clusters whose members are more similar to each other than to those in other clusters \citep{jain1999_clustering}. The dataset was normalized prior to clustering to ensure each feature contributed appropriately. 

In this work, clustering was performed using Lloyd’s algorithm with random initialization, meaning that the final solution was selected as the minimum \emph{inertia} result after the method was initialized 10 independent times. To assess the robustness of the clustering with respect to centroid initialization, the full PCA + K-Means procedure was repeated 100 times using different random seeds. The resulting cluster assignments were compared using the Adjusted Rand Index and Adjusted Mutual Information
\citep{hubert1985,vinh2010}, both of which are invariant to permutations of cluster labels. The clustering was found to be highly stable, yielding mean ARI and AMI values of 0.993 and 0.987, respectively, with minimum values of 0.982 and 0.969 across all pairwise comparisons. These results indicate that the identified cluster structure is largely insensitive to the random initialization of centroids. Inertia is defined as
\begin{equation}
    \text{Inertia} = \sum_{i=1}^{N} \min_{j \in \{1,\dots,k\}} \lVert \mathbf{x}_i - \boldsymbol{\mu}_j \rVert^2
\end{equation}
for $N$ number of data points, $k$ clusters, $x_i$ the $i$-th data point, and $\mu_j$ is the centroid of the $j$-th cluster \citep{lloyd1982}. In summary, inertia is the sum of the squared distances between each sample and its assigned cluster centroid; lower inertia values indicate tighter, more compact clusters.

% The number of clusters was selected by testing multiple values of $k$ and evaluating the resulting cluster stability and physical interpretability. A value of $k=6$ was found to provide the best compromise between capturing distinct solar wind regimes and avoiding over-fragmentation of the dataset. 

% \subsection{Clustering Sensitivity Analysis}\label{methods_sensitivity}
To assess the robustness of the K-Means clustering results, we performed a systematic sensitivity study by varying the number of clusters $k$ in the K-Means algorithm. The analysis was carried out in the six-dimensional PCA-reduced space (PCA$ = 6$), and cluster numbers $k=3,\dots,10$ were tested. Clustering quality was quantified using two complementary metrics: inertia and the silhouette score. The silhouette score measures cluster separation by comparing the mean intra-cluster distance to the mean nearest-cluster distance for each data point; values close to 1 indicate well-separated clusters, whereas values near 0 indicate substantial overlap \citep{ROUSSEEUW198753}. To reduce computational cost for the large dataset, silhouette scores were computed using a random subsample of up to 10,000 points, providing a statistically representative estimate of cluster separation while avoiding the quadratic scaling of the full silhouette calculation. Whilst K-Means is a powerful tool used to identify groupings within the dataset, it is acknowledged that there are alternate methods that would address some of sensitivities that K-means has to outliers for example, which is beyond the scope of this study.

\begin{figure}[htbp]
\centering
\includegraphics[width=\linewidth]{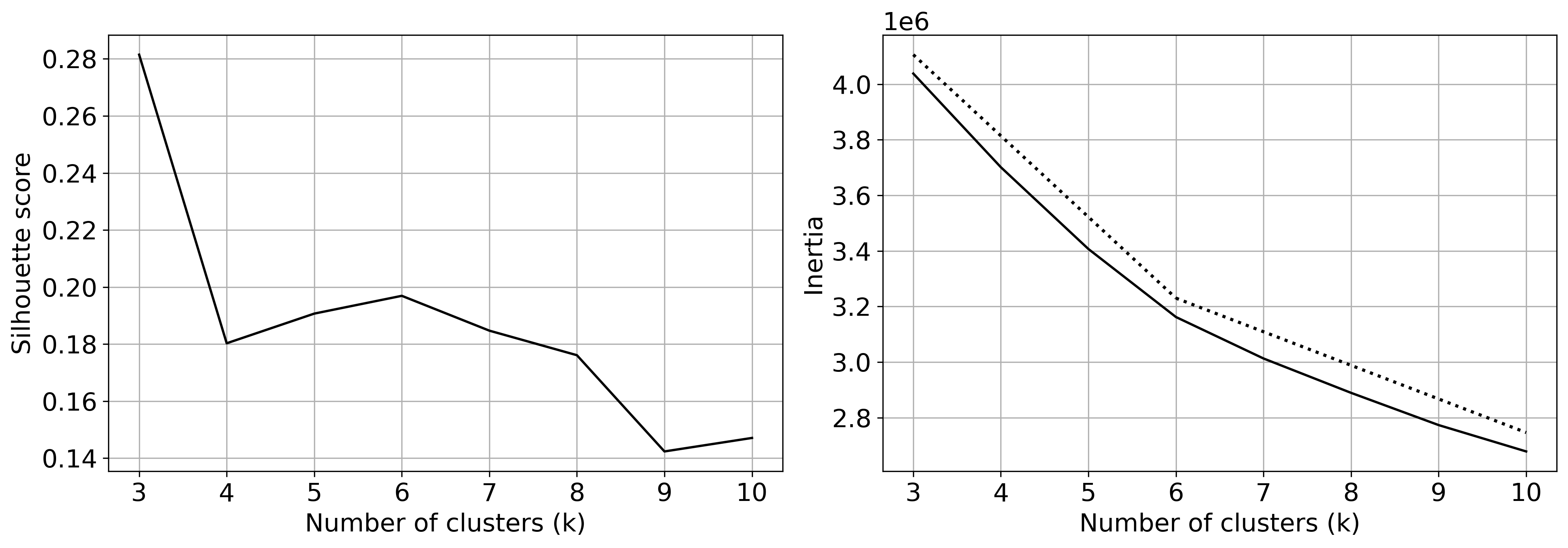}
\caption{
Sensitivity of K-Means clustering results using six retained principal components (PCA = 6).
The left panel shows the silhouette score as a function of the number of clusters ($k$), while the right panel displays the corresponding K-Means inertia.
The dotted line segments in the inertia panel indicate the change in slope around $k=6$, highlighting the elbow that motivates the adopted number of clusters.
}
\label{fig:sensitivity}
\end{figure}

\autoref{fig:sensitivity} presents the clustering sensitivity results for the full dataset spanning Solar Cycles 24-25. The silhouette score (left) exhibits moderate values ($\sim 0.14$--$0.29$) across the explored range of $k$, indicating partial overlap between clusters in the PCA-reduced space. The silhouette score reaches its local maximum near $k=6$, corresponding to a coarse partitioning of the solar wind, and decreases gradually for larger values of $k$, reflecting increasing overlap as the data are subdivided into finer groups.

The inertia (right) decreases monotonically with increasing $k$, as expected, but displays a change in gradient around $k \simeq 6$, highlighted by the dotted guide segments. Beyond this point, further increases in $k$ yield only marginal reductions in inertia, indicating diminishing returns in group compactness. Taken together, these diagnostics suggest that a value of $k=6$ provides a reasonable compromise between maximizing cluster separation at low $k$ and avoiding over-fragmentation at higher $k$, while retaining sufficient granularity to capture distinct solar wind regimes across the full dataset.

Equivalent sensitivity analyses performed for individual solar-cycle phases exhibit similar trends in both inertia and silhouette score, supporting the adopted choice of six clusters as a reasonable compromise across the analyzed temporal subsets.
%%%%%%%%%%%%%%%%%%%%%%%%%%%%%%%%%%%%%%%%%%%%%%%%%%%
\section{Results}\label{results}

\subsection{PCA Structure and Physical Interpretation of K-Means Clusters}\label{sec:phys_interpretation}

Figure 4 presents the percentage contribution of the 14 original features to the first six principal components (PC1–PC6). The heatmap quantifies the percentage of each physical parameter contributing to the variance captured by each principal component, highlighting which variables dominate the reduced-dimensional representation of the data. This is also visualised over time, shown in Figure \ref{fig:clusterdists} in Appendix \ref{app:clusterdists_time}. Overall, the decomposition shows that each principal component is governed by a small subset of features, typically one to three dominant contributors, rather than broad, uniform contributions across all variables. This indicates that the dimensionality reduction preserves physically interpretable structure in the dataset.

\begin{figure}[h]
    \centering
    \includegraphics[width=\linewidth]{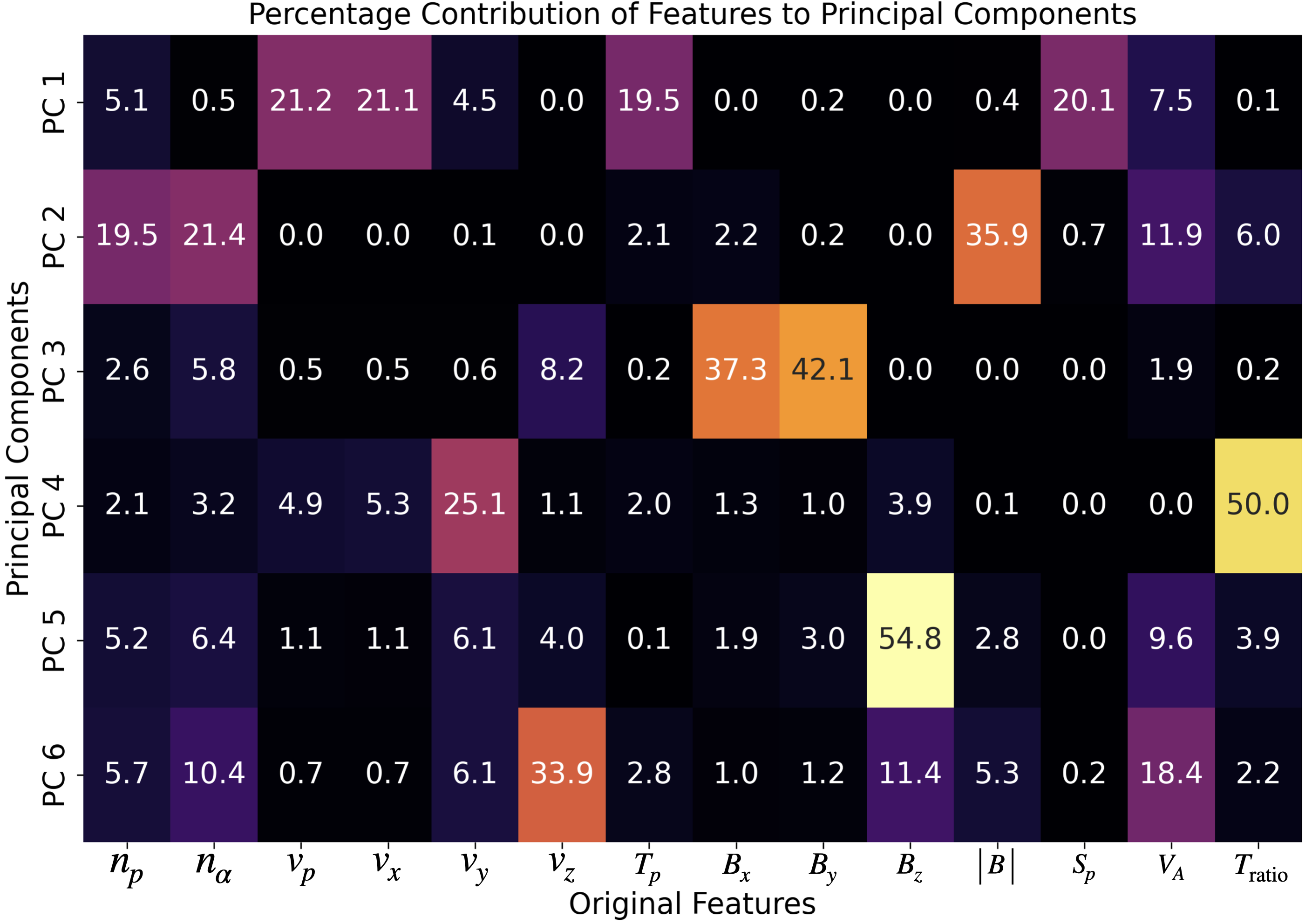}
    \caption{Heatmap showing the percentage contribution of the original features to the first six principal components. The values in the heatmap indicate the percentage contribution of each feature to the respective principal component. This highlights that a majority of each PC is dominated by 1 to 3 original features.}
    \label{fig:PCAloading}
\end{figure}

The leading principal components can be associated with physically interpretable combinations of plasma and magnetic-field variables. 
PC1 is mostly composed of the bulk velocity terms ($v_p$, $v_x$), proton temperature ($T_p$), and proton entropy ratio ($S_p$), indicating that this component primarily captures correlated variations in large-scale flow speed and thermodynamic state. PC2 is governed by proton and alpha particle number densities ($n_p$, $n_\alpha$), magnetic field magnitude ($\left| B \right|$), and Alfvén speed ($V_A$), reflecting changes in plasma density and overall magnetic field strength. PC3 is strongly controlled by the transverse magnetic field components ($B_x$, $B_y$), isolating magnetic field orientation and fluctuation structure. PC4 is dominated by the $y$-component of the velocity ($v_y$) and the temperature ratio ($T_{\mathrm{ratio}}$), suggesting sensitivity to directional flow structure coupled with thermodynamic contrasts. PC5 is overwhelmingly driven by the magnetic field $z$-component ($B_z$), representing dominant variability associated with magnetic field polarity. Finally, PC6 reflects strong contributions from the $z$-component of the velocity ($v_z$), Alfvén speed ($V_A$), and the temperature ratio, indicating coupling between vertical flow structure, magnetic field strength effects, and plasma thermodynamics. The leading principal components separate the dataset into modes corresponding to bulk flow dynamics, density and magnetic field variations, magnetic field geometry, and thermodynamic structure. These physically interpretable modes provide the basis for the clustering analysis, with individual clusters representing distinct combinations of the dominant PCA modes rather than being controlled by any single original variable.

To quantify the physical characteristics of each cluster in terms of the original features, \autoref{tab:cluster_stats_main} reports the mean and standard deviation of selected plasma and magnetic field features computed in the original physical feature space. These averaged values provide a compact and physically interpretable description of the typical solar wind conditions associated with each cluster. Although additional variables contribute strongly to individual principal components (\autoref{fig:PCAloading}), the parameters reported here represent physically integrative quantities that summarize the dominant thermodynamic, flow, and magnetic regimes associated with each cluster. Variables describing directional or component-level structure (e.g., individual velocity or magnetic field components) primarily govern the PCA modes but are less suitable for compact regime-level characterization via mean values, as they often fluctuate around zero and are coordinate dependent.

For completeness, the mean and standard deviation of all physical variables used in the clustering analysis are reported for each cluster in Appendix~\ref{app:cluster_stats_full}.

\begin{table*}[h]
\caption{Mean and standard deviation of selected plasma and magnetic field features for each of the six clusters identified by the K-Means algorithm. Features are grouped into plasma properties (top block) and magnetic/derived properties (bottom block). Here $N$ denotes the number of data points in each cluster.}
\label{tab:cluster_stats_main}
\scriptsize
\begin{tabular}{l c c c}
\hline
\multicolumn{4}{c}{\textbf{Plasma properties}} \\
\hline
Cluster ($N$) &
$n_p$ [cm$^{-3}$] &
$v_p$ [km\,s$^{-1}$] &
$T_p$ [eV] \\
\hline
0 (225\,318) &
$5.01 \pm 2.89$ &
$343 \pm 30$ &
$4.07 \pm 1.38$ \\

1 (61\,360) &
$2.72 \pm 1.39$ &
$550 \pm 49$ &
$16.83 \pm 4.43$ \\

2 (114\,105) &
$2.24 \pm 1.01$ &
$455 \pm 39$ &
$8.36 \pm 2.34$ \\

3 (63\,709) &
$5.45 \pm 2.73$ &
$394 \pm 49$ &
$8.40 \pm 3.86$ \\

4 (67\,427) &
$5.18 \pm 2.53$ &
$398 \pm 43$ &
$8.77 \pm 3.45$ \\

5 (23\,432) &
$16.55 \pm 6.87$ &
$370 \pm 52$ &
$5.57 \pm 2.82$ \\

\hline
\multicolumn{4}{c}{\textbf{Magnetic and derived properties}} \\
\hline
Cluster ($N$) &
$|B|$ [nT] &
$V_A$ [km\,s$^{-1}$] &
$S_p = T_p / n_p^{2/3}$ \\
\hline
0 (225\,318) &
$2.36 \pm 0.85$ &
$24.7 \pm 8.4$ &
$1.68 \pm 0.89$ \\

1 (61\,360) &
$3.50 \pm 1.34$ &
$48.2 \pm 17.8$ &
$9.58 \pm 3.98$ \\

2 (114\,105) &
$2.24 \pm 0.73$ &
$33.6 \pm 9.4$ &
$5.27 \pm 1.77$ \\

3 (63\,709) &
$5.05 \pm 1.66$ &
$49.9 \pm 15.3$ &
$3.12 \pm 1.73$ \\

4 (67\,427) &
$4.38 \pm 1.69$ &
$43.8 \pm 14.1$ &
$3.30 \pm 1.56$ \\

5 (23\,432) &
$6.07 \pm 2.57$ &
$34.5 \pm 16.4$ &
$0.94 \pm 0.57$ \\
\hline
\end{tabular}
\end{table*}

The most populous cluster (Cluster~0) is characterized by relatively low bulk velocity, weak magnetic field strength, low Alfvén speed, and reduced proton temperature, together with moderate proton density. These properties are consistent with classical slow solar wind conditions commonly associated with streamer belt outflows (\cite{Kasper2012,abbo2016}). The dominance of this cluster reflects the prevalence of slow solar wind conditions at Mars over the analyzed time interval.

Clusters~3 and~4 exhibit intermediate bulk velocities and enhanced magnetic field magnitudes relative to Cluster~0, accompanied by moderately elevated proton temperatures and Alfvén speeds. These characteristics suggest transitional solar wind regimes that bridge slow and fast wind conditions. Such clusters likely represent mixed plasma states, including interaction regions between slow and fast streams or evolving large-scale structures within the heliosphere \cite{Richardson2004,Borovsky2012}.

Fast solar wind streams are primarily represented by Clusters~1 and~2. These clusters display elevated bulk velocities, reduced proton densities, increased Alfvén speeds, and significantly higher proton temperatures and entropy proxy values. In particular, Cluster~1 corresponds to the fastest and hottest plasma population, consistent with high-speed streams of coronal hole origin \cite{McComas2008,Cranmer2009}, while Cluster~2 represents a somewhat slower but still fast wind regime with intermediate thermodynamic properties.

Finally, Cluster~5 stands out due to its markedly enhanced proton density, strong magnetic field magnitude, and large variability across multiple parameters, combined with comparatively lower proton temperature and entropy proxy values. This combination of features is indicative of compressed or disturbed plasma conditions, such as stream interaction regions, compressed sheath regions, or other transient solar wind structures \cite{Richardson2018,Borovsky2012}.

The relatively small separation between some cluster-averaged properties highlights the intrinsically continuous nature of solar wind plasma distributions. Consequently, the identified clusters should be interpreted as representative regimes within a continuous parameter space, rather than as sharply distinct or isolated populations.

\subsection{Solar cycles and temporal variability of solar wind regimes}\label{sec:temporal_analysis}

To investigate the influence of large-scale solar activity on the identified solar wind regimes, the dataset was segmented according to different phases of Solar Cycle 24, focusing on representative intervals of solar maximum and solar minimum. Specifically, the solar maximum of Cycle~24 is represented by the interval 1~December~2014 to 15~March~2015, while the solar minimum of Cycle~24 by 1~October~2018 to 15~January~2019. These intervals were selected to balance two considerations: (i) correspondence with established phases of solar activity, and (ii) continuous data coverage, ensuring comparable temporal lengths across all figures.

In the following discussion, cluster numbering follows the physical interpretation established in the previous section.

\begin{figure}[t]
    \centering
    \includegraphics[width=\linewidth]{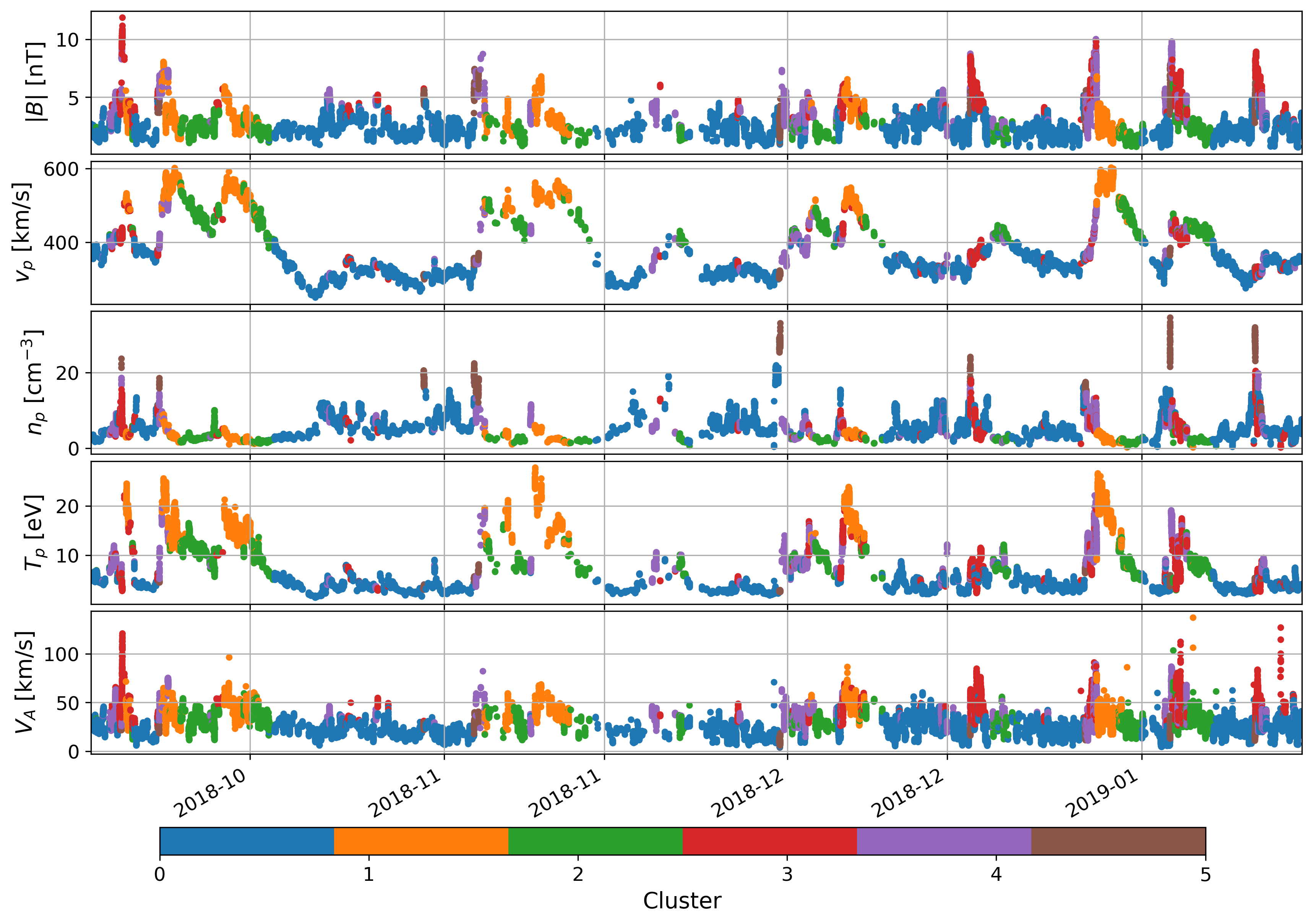}
    \caption{Multi-panel time series of solar wind parameters during the minimum of Solar Cycle~24. Panels show (from top to bottom): magnetic field magnitude $|B|$, proton velocity $v_p$, proton density $n_p$, proton temperature $T_p$, and Alfvén velocity $V_A$. Data correspond to the interval 1~October~2018 -- 15~January~2019. Points are colored by cluster assignment.}
    \label{fig:cycle24_min}
\end{figure}

During the minimum of Solar Cycle~24 (\autoref{fig:cycle24_min}), the time series are dominated by Cluster~0, which appears as extended intervals of relatively low and stable magnetic field strength, bulk velocity, proton density, temperature, and Alfvén speed. This behavior is characteristic of persistent slow solar wind conditions and reflects the reduced dynamical variability typical of solar minimum.

Nevertheless, fast solar wind regimes are not absent. Clusters~1 and~2, together accounting for a non-negligible fraction of the data, appear intermittently as enhancements in proton velocity and Alfvén speed, often embedded within otherwise slow-wind intervals. Occasional short-lived enhancements in magnetic field magnitude and proton density are primarily associated with Cluster~3, indicative of moderately compressed or disturbed plasma structures, while Cluster~5 appears only rarely. Cluster~4 occurs sporadically as a transitional regime, marking gradual transitions between slow and faster wind states. Overall, the predominance of Cluster~0,combined with the relatively low occurrence of disturbed plasma, confirms the comparatively quiescent heliospheric environment during this phase.

\begin{figure}[t]
    \centering
    \includegraphics[width=\linewidth]{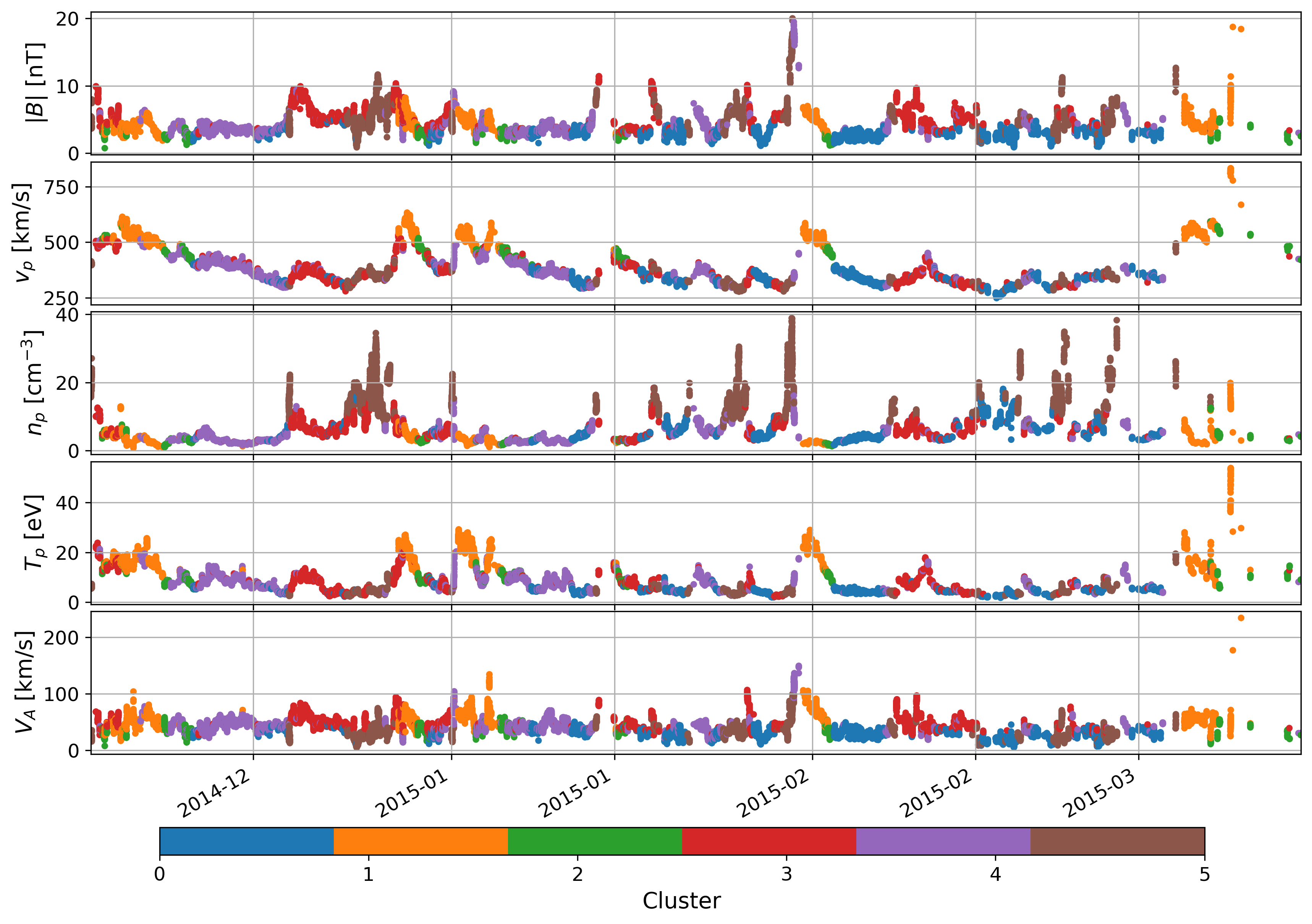}
    \caption{Multi-panel time series of solar wind parameters during the maximum of Solar Cycle~24. Panels show (from top to bottom): magnetic field magnitude $|B|$, proton velocity $v_p$, proton density $n_p$, proton temperature $T_p$, and Alfvén velocity $V_A$. Data correspond to the interval 1~December~2014 -- 15~March~2015. Points are colored by cluster assignment.}
    \label{fig:cycle24_max}
\end{figure}

In contrast, the maximum of Solar Cycle~24 (\autoref{fig:cycle24_max}) exhibits substantially increased variability and a reduced persistence of slow solar wind conditions. Cluster~0 appears less frequently and in shorter intervals, while intermediate and disturbed regimes become more prominent. Cluster~5, corresponding to compressed or disturbed plasma, occurs repeatedly across multiple parameters, often coinciding with enhancements in magnetic field strength and proton density. Extended intervals dominated by Cluster~4, together with recurrent occurrences of Cluster~3, indicate prolonged transitional and moderately disturbed states characterized by moderate bulk velocities and elevated magnetic fields. Fast solar wind streams, represented by Clusters~1 and~2, are intermittently observed, particularly during sharp increases in proton velocity and Alfvén speed. The rapid alternation between cluster regimes highlights the complex and dynamic solar wind structure typical of solar maximum conditions.

Taken together, these results demonstrate that the clusters identified by the unsupervised regime identification correspond to physically interpretable solar wind regimes that recur throughout the analyzed dataset. However, their temporal organization, persistence, and relative occurrence are strongly modulated by solar activity. The solar minimum interval is dominated by long-lived slow solar wind conditions with reduced variability, whereas the solar maximum interval is characterized by frequent regime transitions, enhanced occurrence of intermediate and compressed states, and increased large-scale structuring of the solar wind. This behavior underscores both the robustness of the clustering approach and the central role of solar activity in shaping solar wind variability at Mars.

A qualitative summary of the physical interpretation and typical occurrence of each identified cluster elaborated in \autoref{sec:phys_interpretation} and \ref{sec:temporal_analysis} is provided in Table~\ref{tab:cluster_summary}.

\begin{table}[h]
\centering
\caption{Summary interpretation of the six solar wind regimes identified by the K-Means clustering.}
\label{tab:cluster_summary}
\begin{tabular}{c p{2cm} p{4cm} p{3.5cm}}
\hline
Cluster & Regime label & Dominant characteristics & Typical occurrence \\
\hline
0 & Slow solar wind &
Low bulk velocity, weak magnetic field, low Alfvén speed, cool plasma &
Dominant during solar minimum \\

1 & Fast solar wind (hot) &
Very high speed, low density, high temperature and entropy, elevated $V_A$ &
Solar maximum and coronal hole streams \\

2 & Fast solar wind (moderate) &
High speed, low density, moderate temperature and entropy &
Throughout cycle, often transitional \\

3 & Intermediate / compressed &
Moderate speed, enhanced magnetic field and density &
Stream interaction regions \\

4 & Transitional wind &
Intermediate speed and magnetic field, mixed plasma properties &
Transitions between slow and fast wind \\

5 & Strongly compressed / disturbed &
High density and magnetic field, low entropy, large variability &
Solar maximum, compressed and disturbed intervals \\
\hline
\end{tabular}
\end{table}

%%%%%%%%%%%%%%%%%%%%%%%%%%%%%%%%%%%%%%%%%%%%%%%%%%%

\section{Conclusions}

In this study, we applied an unsupervised machine-learning framework combining PCA and K-Means clustering to a multi-year MAVEN upstream solar wind dataset spanning Solar Cycle~24. By reducing a heterogeneous, multi-parameter plasma dataset to a lower-dimensional representation while preserving its dominant variance structure, we were able to identify recurrent solar wind regimes at Mars. The resulting clusters were examined both in the original physical feature space and through their temporal evolution across different phases of the solar cycle. The physical interpretation of the identified regimes is summarized in Table~\ref{tab:cluster_summary}.

The analysis identifies six clusters that are best interpreted as representative solar wind regimes rather than sharply distinct plasma populations. Cluster~0 corresponds to classical slow solar wind, characterized by low bulk velocity, weak magnetic field strength, reduced Alfvén speed, and comparatively cool plasma, and dominates the dataset, particularly during solar minimum. Fast solar wind conditions are primarily captured by Clusters~1 and~2, which exhibit elevated flow speeds, enhanced Alfvén velocities, higher proton temperatures, and increased entropy proxy values, consistent with high-speed streams of coronal hole origin. Clusters~3 and~4 represent intermediate or transitional regimes, marked by moderate enhancements in magnetic field strength and plasma parameters, and frequently occur during transitions between slow and fast wind states. Cluster~5 isolates strongly compressed and highly variable plasma conditions, characterized by enhanced density and magnetic field magnitude, and is associated with episodic extreme events such as stream interaction regions or compressed sheath-like structures.

The temporal analysis of representative Solar Cycle 24 minimum and maximum intervals, together with the full-dataset occurrence analysis, demonstrates that these regimes recur throughout the MAVEN dataset while their relative occurrence, persistence, and organization vary with solar activity. Solar minimum conditions are dominated by long-lived slow solar wind intervals with reduced variability and infrequent compressed states. In contrast, the representative solar-maximum interval is characterized by more frequent transitions between regimes, an increased occurrence of intermediate and disturbed plasma states, and enhanced temporal variability. In particular, fast and compressed regimes appear more frequently during active phases, reflecting the increased complexity of the heliospheric environment.

This study used similar tools and techniques used in \citep{amaya2020} to classify solar wind conditions, but applied to data at Mars. Their study, using 14 years of ACE data, found that using unsupervised clustering leads to uncertainties in distinguishing characteristic solar wind classes when applied to unprocessed data, and that a combination of feature engineering, non-linear transformations and self-organized map training is needed. Whilst this study uses an already processed solar wind dataset at Mars provided by \cite{Halekas2017a}, this technique could be applied to unprocessed solar wind data collected by MAVEN to compare whether this is also the case at Mars. Cases at Earth and Mars found this technique could distinguish between known heliospherical events and the solar cycle, but at Earth \cite{amaya2020} conclude that more context and processing must occur to make these regime categorizations significant. More recent work has further demonstrated the applicability of unsupervised clustering techniques to identifying physically meaningful solar wind structure in the inner heliosphere \cite{carella2025}. At Mars, we find that the clusters derived from the full MAVEN dataset and from representative solar-cycle intervals provide a useful and physically interpretable framework for investigating solar-wind variability.

Although individual principal components are influenced by multiple physical variables, the combined interpretation of PCA loadings, cluster-averaged plasma properties, and temporal behavior demonstrates that the identified clusters provide a physically meaningful and interpretable partitioning of the solar wind parameter space at Mars. These results highlight the value of unsupervised clustering techniques for characterizing solar wind variability in a data-driven yet physically grounded manner. Alternative clustering methods such as density based, probabilistic, or explainable clustering approaches may reveal additional structure and can be explored in future work. The framework developed here provides a foundation for future comparative studies across heliocentric distances and planetary environments, and for investigating the solar wind drivers of magnetospheric and atmospheric variability at Mars. As continuous solar wind measurements are currently not available at Mars, clusters such as those developed here provide a framework for interpreting available upstream observations and for placing future measurements into a broader solar-wind context. In particular, it offers a natural context for forthcoming multi-point observations by the dual-spacecraft ESCAPADE mission, currently expected to arrive at Mars in 2027.

%%%%%%%%%%%%%%%%%%%%%%%%%%%%%%%%%%%%%%%%%%%%%%%%%%%%%%%%%%%%%%%%%%%%%%%%%%%
%% Appendix
\appendix

\section{Effects of MAVEN Orbital Motion}
MAVEN is a highly elliptical, non-stationary spacecraft whose orbital motion causes it to sample different regions of near-Mars space over time \citep{jakosky2015}. As a result, temporal variations in the observed plasma and magnetic field parameters may reflect not only intrinsic solar wind variability but also changes in spacecraft position relative to Mars, including distance from the bow shock, foreshock regions, and induced magnetosphere boundaries.
While the present analysis does not explicitly correct for spacecraft position, the clustering approach is applied consistently across the full dataset and primarily captures relative variations in plasma properties. Nevertheless, orbital effects may contribute to some of the observed structure in parameter space, particularly during extended time intervals. Future work may incorporate MAVEN ephemeris data to separate spatial sampling effects from intrinsic solar wind variability.

\section{Temporal Evolution of Cluster Occurrence}\label{app:clusterdists_time}

\begin{figure}
    \centering
    \includegraphics[width=\linewidth]{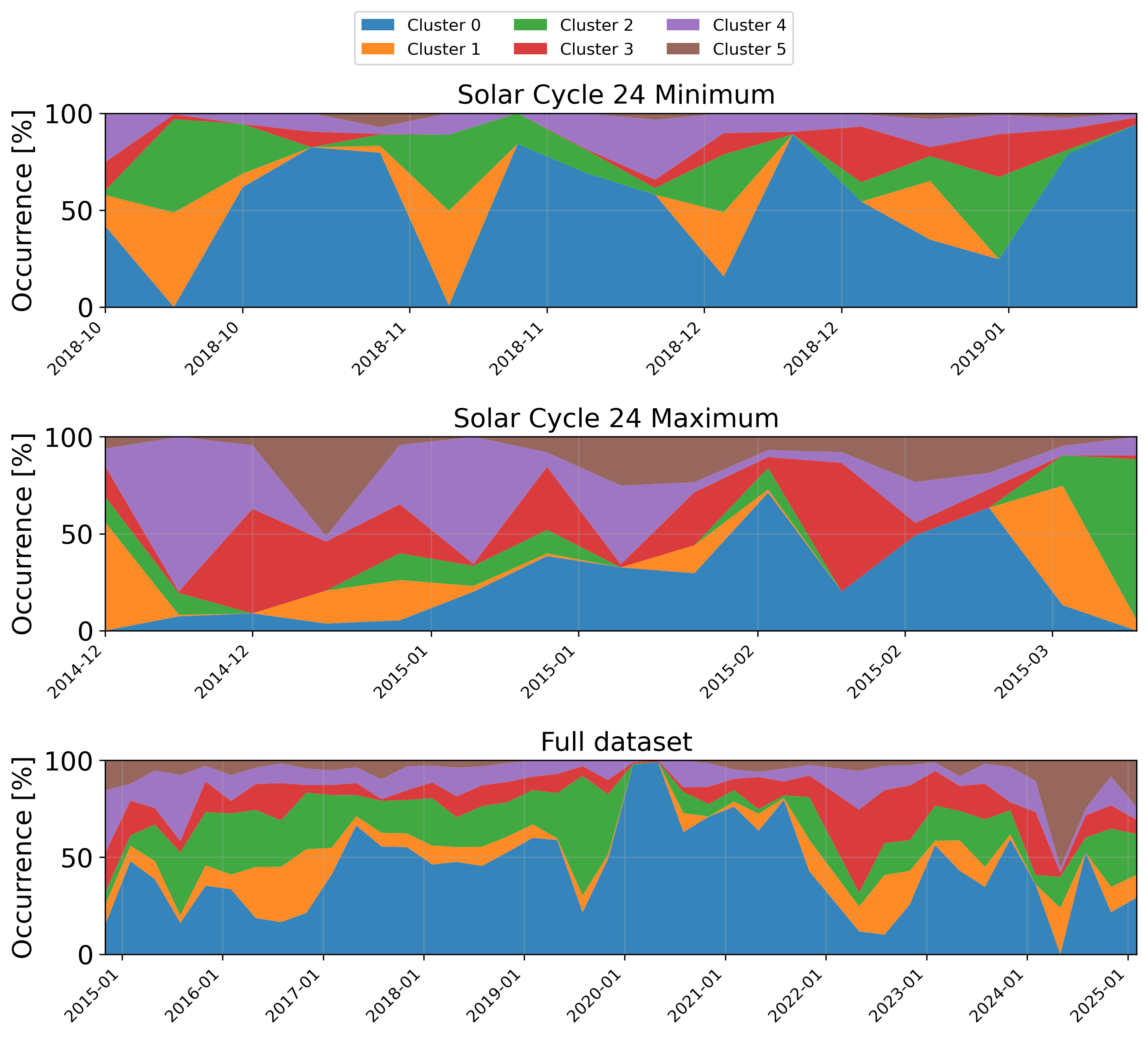}
    \caption{Percentage of observations assigned to each cluster as a function of time.
                The upper two panels show the occurrence rates during representative
                intervals of the solar minimum and solar maximum of Solar Cycle~24,
                while the bottom panel shows the evolution over the full MAVEN dataset
                using three-month bins.}
    \label{fig:clusterdists}
\end{figure}

Figure~\ref{fig:clusterdists} provides additional information on the temporal evolution of cluster occurrence rates. The upper panels correspond to the representative intervals of solar minimum and solar maximum discussed in Section~\ref{sec:temporal_analysis}. During the solar minimum interval, Cluster~0 dominates for extended periods, reflecting the persistence of relatively stable slow solar wind conditions. In contrast, the solar maximum interval exhibits more frequent transitions between clusters and a larger contribution from intermediate and disturbed regimes, consistent with the enhanced variability associated with increased solar activity. \\ The lower panel extends this analysis to the full MAVEN dataset. All identified clusters are present throughout the mission, demonstrating that the clustering captures recurring solar wind regimes rather than transient features specific to a particular epoch. However, their relative occurrence varies substantially over time, with alternating periods of dominance by different clusters. This long-term variability is consistent with the changing heliospheric conditions sampled during different phases of the solar cycle and further supports the robustness of the clustering approach across a wide range of solar wind environments.

\section{Full Cluster Statistics}\label{app:cluster_stats_full}

This appendix reports the mean and standard deviation of the physical variables used in the PCA and K-Means clustering analysis, computed in the original physical feature space for each identified cluster. For completeness, directional velocity and magnetic field components
($v_x$, $v_y$, $v_z$, $B_x$, $B_y$), which fluctuate around zero and are
coordinate dependent, are not tabulated here but are included in the
PCA and clustering analysis.

\begin{table*}[h]
\centering
\caption{Mean $\pm$ standard deviation of plasma parameters for each cluster.}
\label{tab:cluster_stats_plasma}
\scriptsize
\begin{tabular}{l c c c c c}
\hline
Cluster &
$n_p$ [cm$^{-3}$] &
$n_\alpha$ [cm$^{-3}$] &
$v_p$ [km\,s$^{-1}$] &
$T_p$ [eV] &
$S_p = T_p / n_p^{2/3}$ \\
\hline
0 &
$5.01 \pm 2.89$ &
$0.089 \pm 0.077$ &
$343 \pm 30$ &
$4.07 \pm 1.38$ &
$1.68 \pm 0.89$ \\

1 &
$2.72 \pm 1.39$ &
$0.096 \pm 0.061$ &
$550 \pm 49$ &
$16.83 \pm 4.43$ &
$9.58 \pm 3.98$ \\

2 &
$2.24 \pm 1.01$ &
$0.076 \pm 0.061$ &
$455 \pm 39$ &
$8.36 \pm 2.34$ &
$5.27 \pm 1.77$ \\

3 &
$5.45 \pm 2.73$ &
$0.147 \pm 0.110$ &
$394 \pm 49$ &
$8.40 \pm 3.86$ &
$3.12 \pm 1.73$ \\

4 &
$5.18 \pm 2.53$ &
$0.161 \pm 0.104$ &
$398 \pm 43$ &
$8.77 \pm 3.45$ &
$3.30 \pm 1.56$ \\

5 &
$16.55 \pm 6.87$ &
$0.562 \pm 0.424$ &
$370 \pm 52$ &
$5.57 \pm 2.82$ &
$0.94 \pm 0.57$ \\
\hline
\end{tabular}
\end{table*}

\begin{table*}[h]
\centering
\caption{Mean $\pm$ standard deviation of magnetic field and derived parameters for each cluster.}
\label{tab:cluster_stats_magnetic}
\scriptsize
\begin{tabular}{l c c c c c}
\hline
Cluster &
$|B|$ [nT] &
$B_z$ [nT] &
$V_A$ [km\,s$^{-1}$] &
$T_{\rm ratio}$ &
$N$ \\
\hline
0 &
$2.36 \pm 0.85$ &
$0.02 \pm 1.29$ &
$24.7 \pm 8.4$ &
$0.64 \pm 0.15$ &
225\,318 \\

1 &
$3.50 \pm 1.34$ &
$0.28 \pm 1.93$ &
$48.2 \pm 17.8$ &
$0.67 \pm 0.21$ &
61\,360 \\

2 &
$2.24 \pm 0.73$ &
$-0.02 \pm 1.20$ &
$33.6 \pm 9.4$ &
$0.74 \pm 0.19$ &
114\,105 \\

3 &
$5.05 \pm 1.66$ &
$0.11 \pm 2.52$ &
$49.9 \pm 15.3$ &
$0.51 \pm 0.17$ &
63\,709 \\

4 &
$4.38 \pm 1.69$ &
$-0.53 \pm 2.43$ &
$43.8 \pm 14.1$ &
$0.48 \pm 0.15$ &
67\,427 \\

5 &
$6.07 \pm 2.57$ &
$1.16 \pm 3.93$ &
$34.5 \pm 16.4$ &
$0.63 \pm 0.24$ &
23\,432 \\
\hline
\end{tabular}
\end{table*}

\section{Contribution of Features to Principal Components}\label{app:con_feat_prin}

\begin{figure}[h]
    \centering
    \includegraphics[width=0.99\linewidth]{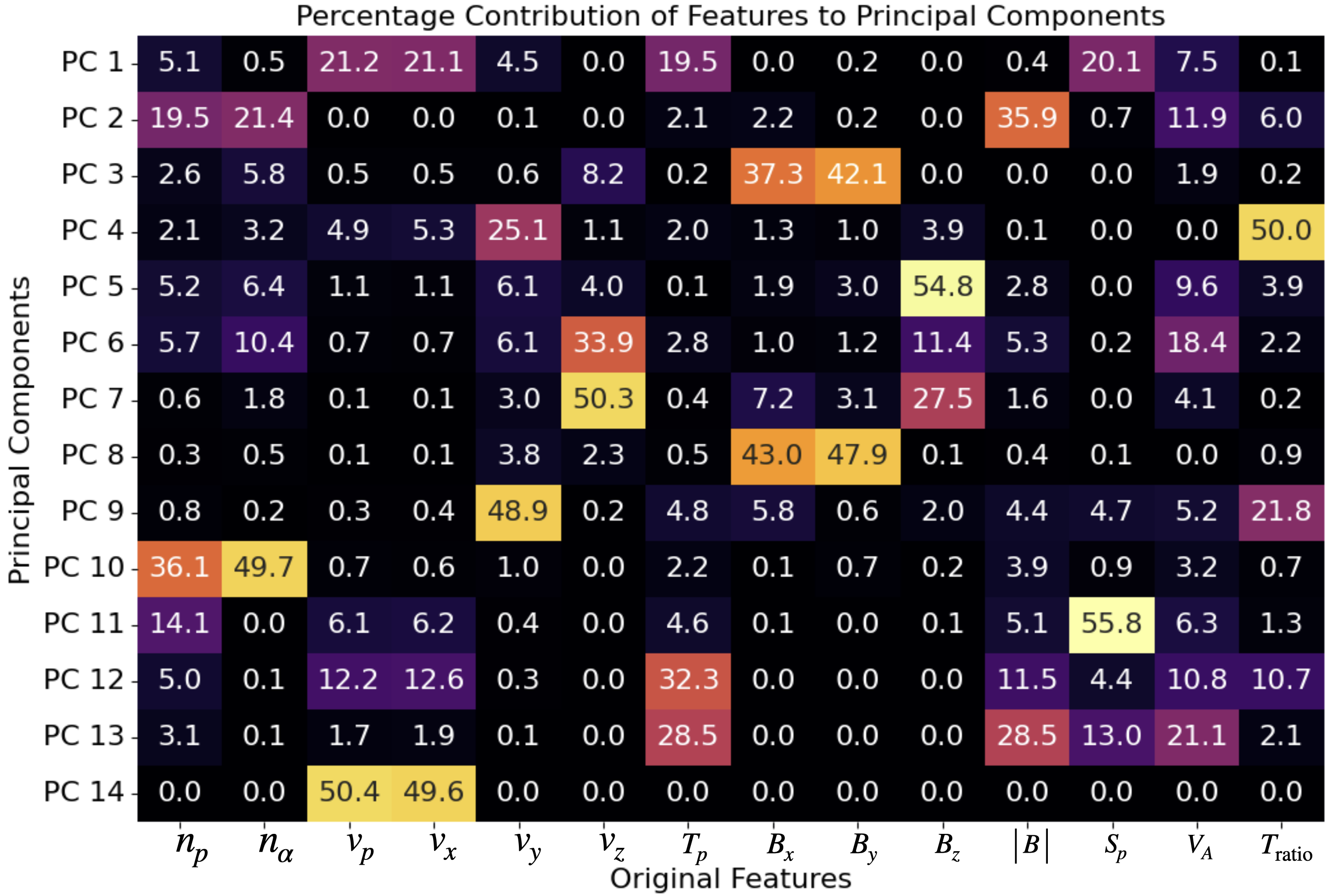}
    \caption{Heatmap showing the percentage contribution of the original features to all of the principal components. The values in the heatmap indicate the percentage contribution of each feature to the respective principal component.}
    \label{fig:confeatpri}
\end{figure}

The clustering analysis presented in this work is based on the first six principal components, which account for the majority of the total variance in the dataset. However, for completeness, the relationship between the 14 features and 14 PCs is shown here in Figure \ref{fig:confeatpri}.

%%%%%%%%%%%%%%%%%%%%%%%%%%%%%%%%%%%%%%%%%%%%%%%%%%%%%%%%%%%%%%%%%%%%%%%%%%%
%% Acknowledgements
%
\begin{acks}
The first three authors, CR, SF and AS all contributed to this paper in equal amounts and should all be considered as first author. This research was conducted during the 19th Heliophysics Summer School 2025 in Boulder, Colorado, a collaboration between NASA and UCAR. The authors would like to acknowledge the support and guidance we received from all faculty and visiting lecturers involved in the school. We also acknowledge Yuanzheng Wen for his involvement during the initial project conception. 
\\ S.F.\ acknowledges support from the FWO PhD fellowship (grant no.\ 1126325N).%I need to have this in every paper I am coauthor during my phd, please do NOT take it off again.
\end{acks}

%% Available additional data environments:
%% required: authorcontribution, fundinginformation, dataavailability
%% optional: materialsavailability, codeavailability
\begin{authorcontribution}
Project conception: CR, SF, AS. Context: CR. Literature review: CR, AA. Data processing and analysis: SF, AS. Analysis and conclusions: CR, SF, AS, AA. Paper review: CR, SF, AS, AA, NG, FK, MV, JH.
\end{authorcontribution}

\begin{fundinginformation}
This research is supported by the NASA Heliophysics: Living With a Star Strategic Capabilities administered by UCAR's Cooperative Programs for the Advancement of Earth System Science (CPAESS) under award 80NSSC22M0097.
\end{fundinginformation}

\begin{dataavailability}
The initial MAVEN solar wind dataset used in this study is publicly available at \url{https://homepage.physics.uiowa.edu/~jhalekas/drivers.html}. 
The analysis code used to process the data, perform PCA, and apply K-Means clustering is available upon reasonable request from the corresponding author.
\end{dataavailability}
%
% \begin{ethics}
% \begin{conflict}
%
% \end{conflict}
% \end{ethics}

%%% %%%%%%%%%%%%%%%%%%%%%%%%%%%%%%%%%%%%%%%%%%%%%%%%%%%%%%%%%%%
%% Bibliography
%
% Using BibTeX
%
\bibliographystyle{spr-mp-sola}
\bibliography{bibliography}  
%
% Without BibTeX 
% \begin{thebibliography}{}
% \bibitem[\protect\citeauthoryear{Author}{Year}]{key}
%   <bibliographical entry>
%
% \bibitem[\protect\citeauthoryear{}{}]{}
%   
%  
% \end{thebibliography}

\end{document}